\documentclass[12pt]{article}

\def\thefootnote{\fnsymbol{footnote}}

\newcommand {\ee}{\end{equation}}
\newcommand {\bea}{\begin{eqnarray}}
\newcommand {\eea}{\end{eqnarray}}
\newcommand {\nn}{\nonumber \\}

\newcommand {\tr}{{\rm tr\,}}

\newcommand {\pl}{\partial}

\newcommand {\vp}{\varphi}

\newcommand {\al}{\alpha}
\newcommand {\be}{\beta}
\newcommand {\ga}{\gamma}

\newcommand {\la}{\lambda}

\newcommand {\si}{\sigma}

\newcommand {\del}  {\delta}

\newcommand {\ab}   {{\alpha\beta}}

\newcommand {\half}{ {\frac{1}{2}} }

\newcommand {\sql} {\sqrt{l}}

\newcommand {\Lcal}{{\cal L}}

\newcommand {\Ncal}{{\cal N}}

\def\overleftarrow#1{\vbox{\ialign{##\crcr
 $\leftarrow$\crcr\noalign{\kern-1pt\nointerlineskip}
 $\hfil\displaystyle{#1}\hfil$\crcr}}}

\newcommand {\labar}{{\bar \lambda}}

\newcommand {\sibar}{{\bar \sigma}}

\newcommand {\alp}{{\alpha'}}
\newcommand {\bep}{{\beta'}}
\newcommand {\gap}{{\gamma'}}

\newcommand  {\xf}{{x^5}}

\newcommand {\ra} {\rightarrow}

\newcommand {\com}  {{\quad ,}}
\newcommand {\q}    {\quad}

\usepackage{graphicx}
\usepackage{latexsym}

\begin{document}
\begin{flushright}
March 2004\\
US-03-07\\
\end{flushright}

\vspace{0.5cm}

\begin{center}

{\Large\bf 
A Bulk Effect to SUSY Effective Potential\\
in a 5D Super-Yang-Mills Model}

\vspace{1.5cm}
{\large Shoichi ICHINOSE
         \footnote{
E-mail address:\ ichinose@u-shizuoka-ken.ac.jp
                  }
}\ and\ 
{\large Akihiro MURAYAMA$^\ddag$
         \footnote{
E-mail address:\ edamura@ipc.shizuoka.ac.jp
                  }
}
\vspace{1cm}

{\large 
Laboratory of Physics, 
School of Food and Nutritional Sciences, 
University of Shizuoka, 
Yada 52-1, Shizuoka 422-8526, Japan
 }

$\mbox{}^\ddag${\large
Department of Physics, Faculty of Education, Shizuoka University,
Shizuoka 422-8529, Japan
}
\end{center}

\vspace{2cm}

\begin{center}
{\large Abstract}
\end{center}

Supersymmetric effective potential of a 5D super-Yang-Mills model compactified on $S^1/Z_2$, i.e., on an   
interval $l$ of extra dimension, is estimated at the 1-loop level by the auxiliary field tadpole
method. For the sake of infinite towers of Kaluza-Klein excitation modes of bulk fields involved in the tadpoles, there
arises a definite bulk effect of linear growth of the effective potential along with the cutoff ${\mit\Lambda}$ which 
is greatly suppressed by $l$ to produce a finite contribution. Incorporating the tree potential and a
Fayet-Iliopoulos $D$-term, the effective potential is minimized at a specific value of $l$, corresponding to an
intermediate mass scale $10^{11-14}$ GeV, where the supersymmetry is restored. 

\newpage
\renewcommand{\thefootnote}{\arabic{footnote}}
\setcounter{footnote}{0}

\noindent {{\bf 1}\q{\bf Introduction}}\\

Recently theories of extra-dimensions have attracted attention. Among them a 5-dimensional (5D) super-Yang-Mills
(super-YM) theory with mirror-plane boundaries is very interesting since it has a possibility to lead to a realistic model
of particle theory. In a previous paper \cite{Ichinose1}, we have analyzed the background configuration based on the
Mirabelli-Peskin \cite{Mirab}-Hebecker \cite{Heb} model and obtained the 1-loop effective potential for some special
cases in a 5D bulk-boundary theory compactified on $S^1/Z_2$ orbifold, i.e., on an interval of length $l$. One of
virtues of the model is that the coupling of a 5D super-YM multiplet to a 4D orientifold  boundary is explicitly given in an
off-shell formulation.

In this paper we try to evaluate a full 1-loop effective potential of 4D boundary in the same framework as
\cite{Ichinose1} except adding a superpotential in the boundary. We use the  auxiliary field tadpole method (AFTM) by
Miller \cite{Miller} based on the tadpole method by S. Weinberg \cite{Weinb}, without eliminating auxiliary  fields by
their equation of motion. An advantage of this method is that only $F$ and $D$ auxiliary field tadpoles are sufficient to
reconstruct the effective potential since the spin-0 tadpole cotributions are generated automatically by the use of a
supersymmetric (SUSY) boundary condition. 

The model is generically non-renormalizable and should be viewed as an effective theory valid up to some high mass
scale associated with an ultraviolet cutoff ${\mit\Lambda}$. However, we require that it should be
renormalizable in the limit of $l \rightarrow 0$.

Although the effective potential to be evaluated is 4D, we have a
definite {\it bulk effect} which comes from the contribution of whole Kaluza-Klein (KK) excitation modes of the
bulk fields involved in the tadpole diagrams. Such a bulk effect is very interesting since it might
implement new aspects of breakings of gauge symmetry and/or supersymmetry through the minimization of
the effective potential. In particular, by minimizing the effective potential which contains the tree level
potential together with an additional contribution of Fayet-Iliopoulos (FI) $D$-term, we find a case that SUSY is
restored at a specific value of the radius $l$ of extra dimension corresponding to an intermediate mass scale $\approx 3
\times 10^{11}$ or $7 \times 10^{13}$ GeV for the ultraviolet cutoff ${\mit \Lambda} \approx M_{GUT}$ or $M_{Pl}$.
\\

\noindent{{\bf 2}\q {\bf 5D super-Yang-Mills model}}\\

Let us consider the 5D flat space-time with the signature $(+----)$. The space of the fifth component is
taken to be $S^1$ with the periodicity $2l$ and the $Z_2$-orbifold condition $\xf \sim -\xf$.
We take a 5D SUSY action such as
\begin{eqnarray}
S=\int d^5X\{\Lcal_{blk}+\del(x^5)\Lcal_{bnd}+\del(x^5-l){\Lcal'}_{bnd}
\},
\label{mp1}
\end{eqnarray}
where $X\equiv (x^0, x^1, x^2, x^3, x^5)$, $\displaystyle{\int dX^5\equiv \int d^4x\int_{-l}^l dx^5}$, 
$\Lcal_{blk}$ is a 5D bulk Lagrangian and $\Lcal_{bnd}$ and $\Lcal'_{bnd}$ denote a 4D boundary
Lagrangian on a "wall" at $\xf=0$ and a hidden sector Lagrangian on the other "wall" at $\xf=l$,
respectively.

The bulk dynamics is given by the 5D super-YM theory which is made of a vector field $A^M\ (M=0,1,2,3,5)$, 
a scalar field $\mit\Phi$, a doublet of symplectic Majorana fields $\la^i\ (i=1,2)$, 
and a triplet of auxiliary scalar fields $X^a\ (a=1,2,3)$:
\begin{eqnarray}
\Lcal_{blk}=&-&\half\tr ({F_{MN}})^2+\tr (\nabla_M\mit\Phi)^2 \nn
& + & \tr (i{\bar\la}^i\gamma^M \nabla_M\lambda^i) 
+ \tr (X^a)^2-\tr ({\bar \la}^i [\mit\Phi,\la^i]),   
\label{mp2}
\end{eqnarray}
where all bulk fields are of the adjoint representation of the gauge group $G: A^M = A^{M\al}T^\al$,
etc., $\tr[T^\al T^\be]= \del^{\al\be}/2$ and $\nabla_M\mit\Phi = \pl_M\mit\Phi - ig[A_M,\mit\Phi]$. 
This system has the symmetry of 8 real supercharges.

We can project out $\Ncal=1$ SUSY multiplet, which has 4 real super charges, 
by assigning $Z_2$-parity to all fields in accordance with the 5D SUSY. 
A consistent choice is given as:\  $P=+1$ for $A_m ($m=0,1,2,3$), \la_L, X^3$;  $P=-1$ for 
$A_5, {\mit\Phi}, \la_R, X^1, X^2$. (The fields of $P=-1$ vanish on the boundaries $x^5 = 0, l$.) Then,
$V\equiv(A_m,\la_L,X^3-\nabla_5{\mit\Phi})$ and
${\mit\Sigma}\equiv({\mit\Phi}+iA_5,-i\sqrt{2}\la_R,X^1+iX^2)$ constitute an $\Ncal =1$ vector
supermultiplet in Wess-Zumino gauge and a chiral scalar supermultiplet, respectively.  Especially
$X^3-\nabla_5{\mit\Phi}\equiv D^{(5)}$ plays the role of $D$-field 
on the wall, namely $D^{(5)}|_{x^5=0,l} = X^3 - \pl_5{\mit\Phi} \equiv
(2l)^{-\half}D$.\footnote{It looks that $D \rightarrow 0$ as $l \rightarrow 0$ at first sight. However, if we
introduce a dimensionless effective 4D gauge coupling, ${\hat g}^2\equiv g^2/(2l)$ which is fixed for $l
\rightarrow 0$, we have $gD^{(5)}|_{x^5=0}={\hat g}D$ irrespective of $l$.}  

We introduce a 4D chiral supermultiplet\footnote{We do not introduce extra 5D matter multiplets (the hypermultiplets)
differently from \cite{Heb}.} $S \equiv (\phi,\psi,F)$ of the fundamental representation which is localized on the wall,
where $\phi,\psi$ and $F$ stand for a complex scalar field , a Weyl spinor and an auxiliary field of complex scalar,
respectively.  This is the simplest matter content on the wall. Using the $\Ncal=1$ SUSY property, we can find the
following  boundary Lagrangian with a definite supersymmetric coupling between bulk and boundary fields:
\begin{eqnarray}
\Lcal_{bnd}&=&S^\dag e^{gV}S\vert_{\theta^2{\bar \theta}^2}+ W(S)\vert_{\theta^2}\nn
&=&\nabla_m\phi^\dag \nabla^m\phi+\psi^\dag i\sibar^m\nabla_m \psi+F^\dag F 
\!-\!\sqrt{2}g(\phi^\dag\la^t_L\si^2\psi+\psi^\dag \si^2\la^*_L\phi)
\!+\! g\phi^\dag D^{(5)}\phi\nn
&& - [m_{\alp\bep}(\phi_\alp F_\bep - \half \psi_\alp\psi_\bep) + \half
\lambda_{\alp\bep\gap}(\phi_\alp\phi_\bep F_\gap - \psi_\alp\psi_\bep\phi_\gap) +\mbox{h.c.}],
\label{mp7}
\end{eqnarray}
where $\nabla_m\equiv \pl_m-igA_m$, $\alp,\bep$ and $\gap$are the suffices of the fundamental representation and
we have taken the following superpotential:
\begin{equation}
W(S) = \half m_{\alp\bep}S_\alp S_\bep + \frac{\lambda_{\alp\bep\gap}}{3!}S_\alp S_\bep
S_\gap,\label{eq33}
\end{equation}
with the coefficients $m_{\alp\bep}$ and
$\lambda_{\alp\bep\gap}$ being such that the gauge symmetry is respected.

Since the hidden sector is irrelevant to the present purpose, we do not specify its Lagrangian $\Lcal_{bnd}'$.\\

\noindent{{\bf 3}\q {\bf Effective Lagrangian for AFTM}}\\

The 1-loop SUSY effective potential $V_{1-loop}$ can be calculated
only by the scalar loop (tadpole)  up to the $F$- and $D$-independent terms
in the off-shell treatment in which the auxiliary fields $F$ and $D$ are not eliminated by their equations of
motion. This is because the auxiliary fields cannot have the Yukawa coupling with fermions and vectors.
This method is called "auxiliary field tadpole method (AFTM)"\cite{Miller}. 

The evaluation of the 1-loop effective potential $V_{1-loop}$ according to AFTM is by the following
recipe:

(1) Find an effective Lagrangian by translating auxiliary and spin-0
fields such that "original field" $\to$ "classical part (VEV)" + "quantum part" .

(2) Write an effective action of the translated theory with the effective Lagrangian plus the source terms, set aside
all tems quadratic in the quantum fields to get ${\cal L}^{(2)}$ and calculate from the generating functional {\it full}
propagators of those which couple with the auxiliary fields.

(3) Evaluate a 1PI 1-point vertex function ${\mit\Gamma}^{(1)}$ for the relevant tadpole diagrams and its
momentum space representation, up to the delta function for the momentum conservation, which is
nothing but the 1-loop auxiliary field tadpole amplitude ${\hat {\mit\Gamma}}_{p_{ext}=0}^{(1)}$ in the
momentum space at zero external momentum.

(4) Integrate the equation
\begin{equation}
\frac{\partial V_{1-loop}}{\partial \langle\mbox{auxiliary field}\rangle }=-{\hat {\mit\Gamma}}_{p_{ext}=0}^{(1)},
\end{equation}
to obtain $V_{1-loop}$, where $\langle\cdots\rangle $ means the VEV.

(5) Determine the final form of $V_{1-loop}$ by making use of SUSY boundary
condition, i.e., $V_{1-loop}(\langle\mbox{auxiliary field}\rangle =0)=0$.

To begin with, we put the following conditions: 
\begin{eqnarray}
A_m=0\ (m=0,1,2,3)\com\q \la^i=\labar^i=0
\com\q \psi=0. \label{ep1}
\end{eqnarray}
to secure the scalar property of the vacuum. The extra (fifth) component of the bulk vector $A_5$ is not
taken to be zero because it is regarded as a 4D scalar on the wall. 

Then, we split all the scalar fields (${\mit\Phi}, X^3, A_5; \phi, F$) into the {\it quantum field} (which is denoted again
by the same symbol) and the {\it classical field} (VEV) ($\vp \equiv \langle{\mit\Phi}\rangle , \chi^3 \equiv \langle 
X^3\rangle , a_5 \equiv \langle A_5\rangle ; \eta \equiv \langle\phi\rangle , f \equiv \langle F\rangle $) as follows:
\begin{eqnarray}
{\mit\Phi}\ra\vp+{\mit\Phi}, X^3\ra \chi^3+X^3,  A_5\ra a_5+A_5,  \phi\ra\eta+\phi,  F\ra f + F.
\label{ep5}
\end{eqnarray}
We allow the classical part of bulk fields $\vp, \chi^3, a_5$ to depend in general on the extra
coordinate
$x^5$. These VEVs do not violate the $Z_2$ symmetry as far as they obey the boundary condition.

The quadratic part of action which is relevant for the present purpose is given by
\begin{eqnarray}
S^{(2)}[{\mit\Phi}, \!\!\!\!\!\!\!\!&,&\!\!\!\!  A_5;\phi, F] = \int d^5X
[\Lcal_{blk}^{(2)}+\del(x^5)\Lcal^{(2)}_{bnd} + \mbox{source terms}],
\label{mp4}\\
\Lcal^{(2)}_{blk} & = & \half\pl_M{\mit\Phi}_\al\pl^M{\mit\Phi}_\al
+\half\pl_M A_{5\al}\pl^M A_{5\al} \nn
& - & gf_{\ab\ga}\{\pl_5\vp_\al A_{5\be}{\mit\Phi}_\ga 
+\pl_5{\mit \Phi}_\al (a_{5\be}{\mit\Phi}_\ga+ A_{5\be}\vp_\ga) \} \nn
& - & g^2f_{\ab\tau}f_{\ga\del\tau}a_{5\al}\vp_\be A_{5\ga}{\mit\Phi}_\del
-\frac{g^2}{2}\{f_{\ab\tau}(a_{5\al}{\mit \Phi}_\be+A_{5\al}\vp_\be)  \}^2, \\
\Lcal^{(2)}_{bnd} & = & \pl_m\phi^\dag\pl^m\phi + {\hat g}\{{\hat d}_\al\,\phi^\dag T^\al\phi -
\pl_5{\mit\Phi}_\al(\eta^\dag T^\al\phi+\phi^\dag T^\al\eta)\} + F^\dag F \nn 
&& 
-\frac{{\hat g}^2}{2}\del(0)(\eta^\dag T^\al\phi+\phi^\dag T^\al\eta)^2 \nn 
&& - [\phi_\alp(m_{\alp\bep} + \lambda_{\alp\gap\bep}\eta_\gap)F_\bep +
\half\lambda_{\alp\bep\gap}\phi_\alp\phi_\bep f_\gap + \mbox{h.c.}],
\label{ep6}
\end{eqnarray}
where ${\hat d}_\al \equiv \langle D_\al\rangle = (2l)^{\half}(\chi_\al^3 - \pl_5\vp_\al), \q
\phi^\dag T^\al\phi\equiv \phi^\dag_\alp(T^\al)_{\alp\bep}\phi_\bep,$ etc. and the 5D auxiliary field
$X^3$ has been integrated out at the price of giving rise to a singular term ($\propto \del(0)$)\cite{Mirab}. \\

\noindent{{\bf 4}\q {\bf Mass-Matrix and the 1PI vertex function}}\\

We are now ready for the calculation of the 1-loop effective potential. 

The effective action (\ref{mp4}) can be expressed as
\begin{eqnarray}
S^{(2)} = \int d^5X \left[\half {\mit\Psi}^{\dagger}M{\mit\Psi}+{\mit\Psi}^{\dagger}J\right], \label{eqn:12}
\end{eqnarray}
where
\begin{eqnarray}
{\mit\Psi}_A^{\dagger} & = & \left(
\phi^\dag_{\al'},\q \phi_{\al'}^t,\q F^\dag_\alp,\q F_\alp^t,\q {\mit\Phi}_\al^t,\q A_{5\al}^t \right),\label{eqn:13}\\
J_A^t & = & \left(
 J_{\phi^\dag_{\al'}}^t,\q J_{\phi_{\al'}}^t,\q J_{F^\dag_\alp}^t,\q J_{F_\alp}^t,\q J_{{\mit\Phi}_\al}^t,\q
J_{A_{5\al}}^t \right) ,
\end{eqnarray}
with $A=(\al', \al), B=(\be', \be)$ and $J'$s denote sources. 

We can perform the integration of (\ref{eqn:12}) w.r.t. $x^5$ by KK-expanding ${\mit\Phi}$ and $A_5$ as follows;
\begin{eqnarray}
{\mit\Phi}_\al(x,x^5) & = & \frac{1}{\sql}\sum_{n=1}^\infty{\mit\Phi}_{n\al}(x)\sin(\frac{n\pi}{l}x^5), \\
A_{5\al}(x,x^5) & = & \frac{1}{\sql}\sum_{n=1}^\infty A_{n\al}(x)\sin(\frac{n\pi}{l}x^5).
\end{eqnarray} 
We obtain
\begin{eqnarray}
S^{(2)} = \int d^4x \left[\half {\hat{\mit\Psi}}^{\dagger}{\cal M}{\hat{\mit\Psi}} + {\hat{\mit\Psi}}^{\dagger}{\hat
J}\right],
\end{eqnarray}
where
\begin{eqnarray}
{\hat{\mit\Psi}}^{\dagger} & = & \left(
\phi^\dag_{\al'},\q \phi_{\al'}^t,\q F^\dag_\alp,\q F_\alp^t,\q {\hat{\mit\Phi}}_\al^t,\q {\hat A}_{5\al}^t \right), \\
{\hat J}_A^t & = & \left(
 J_{\phi^\dag_{\al'}}^t,\q J_{\phi_{\al'}}^t,\q J_{F^\dag_\alp}^t,\q J_{F_\alp}^t,\q {\hat J}_{{\mit\Phi}_\al}^t,\q
{\hat J}_{A_{5\al}}^t \right) ,
\end{eqnarray}
with
\begin{eqnarray}
{\hat{\mit\Phi}}_\al^t & = & ({\mit\Phi}_{1\al},\q {\mit\Phi}_{2\al},\q  \cdots), \\
{\hat A}_\al^t & = & (A_{1\al},\q A_{2\al},\q\cdots), \\
{\hat J}_{{\mit\Phi}_\al}^t & = & (J_{{\mit\Phi}_{1\al}},\q J_{{\mit\Phi}_{2\al}},\q  \cdots), \\
{\hat J}_{A_{5\al}}^t & = & (J_{A_{1\al}},\q J_{A_{2\al}},\q  \cdots), 
\end{eqnarray}
and
\begin{eqnarray}
({\cal M}_{AB}) & = & \left(\begin{array}{ccc}
		{\cal A}_{\alp\bep} & {\cal B}_{\alp\be} & 0 \\
		{\cal C}_{\al\bep} & {\cal M}_{{\hat{\mit\Phi}}_\al{\hat{\mit\Phi}}_\be} & {\cal M}_{{\hat{\mit\Phi}}_\al{\hat
A}_{5\be}} \\
		0 & {\cal M}_{{\hat A}_{5\al}{\hat{\mit\Phi}}_\be} & {\cal M}_{{\hat A}_{5\al}{\hat A}_{5\be}}
\end{array}\right), \label{matrix10}\\
{\cal A}_{\alp\bep} & = & \left(\begin{array}{cccc}
		{\cal M}_{\phi^\dag\phi} & {\cal M}_{\phi^\dag\phi^\dag} & 0 & {\cal M}_{\phi^\dag F^\dag} \\
		{\cal M}_{\phi\phi} & {\cal M}_{\phi\phi^\dag} & {\cal M}_{\phi F} & 0 \\
		0 & {\cal M}_{F^\dag \phi^\dag} & I & 0 \\
		{\cal M}_{F \phi} & 0 & 0 & I 
\end{array}
\right)_{\alp\bep}, \\ 
{\cal B}_{\alp\be} & = & \left(\begin{array}{c}
{\cal M}_{\phi^\dag{\hat{\mit\Phi}}} \\
{\cal M}_{\phi{\hat{\mit\Phi}}} \\ 
0 \\
0 \\
\end{array}\right)_{\alp\be}, \\
{\cal C}_{\al\bep} & = & \left(\begin{array}{cccc}
		{\cal M}_{{\hat{\mit\Phi}}\phi},  & {\cal M}_{{\hat{\mit\Phi}}\phi^\dag},  & 0, & 0 
\end{array}\right)_{\al\bep},
\end{eqnarray}
with
\begin{eqnarray}
{\cal M}_{\phi_\alp^\dag\phi_\bep} &=& -\Box\del_{\alp\bep}+{\hat g}{\hat d}_\ga (T^\ga)_{\alp\bep}
		-g^2\del(0)(T^\ga\eta)_\alp (\eta^\dag T^\ga)_\bep,\nn
{\cal M}_{\phi_\alp\phi_\bep^\dag} &=& -\Box\del_{\alp\bep}+ {\hat g}{\hat d}_\ga(T^\ga)_{\bep\alp}
		-g^2\del(0)(\eta^\dag T^\ga)_\alp(T^\ga\eta)_\bep,\nn  
{\cal M}_{\phi_\alp^\dag\phi_\bep^\dag} &=& - \lambda^*_{\alp\bep\gap}f^\dag_\gap
		+g^2\del(0)(T^\ga\eta)_\alp (T^\ga\eta)_\bep, \nn
{\cal M}_{\phi_\alp\phi_\bep} &=& - \lambda_{\alp\bep\gap}f_\gap +g^2\del(0) (\eta^\dag T^\ga)_\alp (\eta^\dag
		T^\ga)_\bep, \nn 
{\cal M}_{F_\alp^\dag\phi_\bep^\dag} & = & ({\cal M}_{\phi^\dag F^\dag})_{\alp\bep} = -(m^*_{\alp\bep} +
		\lambda^*_{\alp\gap\bep}\eta^\dag_\gap) \equiv \chi^\dag_{\alp\bep}, \nn
{\cal M}_{F_\alp\phi_\bep} & = & ({\cal M}_{\phi F})_{\alp\bep} = - (m_{\alp\bep}+
		\lambda_{\alp\gap\bep}\eta_\gap) \equiv \chi_{\alp\bep}, \nn  
{\cal M}_{\phi_\alp^\dag{\mit\Phi}_{n\be}} &=&-g(T^{\be}\eta)_{\alp}(n\pi/l^{\frac{3}{2}}),\nn 
{\cal M}_{{\mit\Phi}_{n\al}\phi_\bep^\dag}&=&-g(T^\al\eta)_\bep(n\pi/l^{\frac{3}{2}}), \nn  
{\cal M}_{\phi_\alp{\mit\Phi}_{n\be}} &=& -g(\eta^\dag T^\be)_\alp(n\pi/l^{\frac{3}{2}}), \nn  
{\cal M}_{{\mit\Phi}_{n\al}\phi_\bep} &=&-g(\eta^\dag T^\al)_\bep(n\pi/l^{\frac{3}{2}}), \nn 
{\cal M}_{{\mit\Phi}_{m\al}{\mit\Phi}_{n\be}} &=& - (\Box + (n\pi/l)^2)\del_{mn}\del_{\al\be}. \label{calM}
\end{eqnarray}
The explicit form of sources ${\hat J}$'s is not required in the following computation. The matrix elements
$(M_{{\hat{\mit\Phi}}{\hat A}_5})_{\al\be}, (M_{{\hat A}_5{\hat{\mit\Phi}}})_{\al\be}$ and $(M_{{\hat
A}_5A_5})_{\al\be}$ do not depend on $f_\alp, f^\dag_\alp$ and ${\hat d}_\al$ so that they are irrelevant  to the
effective potential to be estimated.

The generating functional $Z[{\hat J}]$ is given by
\begin{equation}
\ln Z[{\hat J}]=-{\frac12}\int d^4xd^4y {\hat J}^\dagger (x) i{\cal M}^{-1}(x,y){\hat J}(y),
\end{equation}
from which we can extract a {\it full} propagator
${\it \Delta}_F(x-y)_{ij}$ through $\delta^2\ln Z[{\hat J}]/\delta {\hat J}_j\delta {\hat J}_i^\dagger$, namely
\begin{equation}
{\it \Delta}_F(x-y)_{ij}=-{\cal M}^{-1}_{ij}=-\frac{m_{ji}}{\mbox{det}{\cal M}},
\end{equation}
where $m_{ji}$ denotes the $j$-$i$th minor of ${\cal M}$. 

The 1PI vertex function ${\mit\Gamma}^{(1)}$ corresponding to the auxiliary field tadpole is defined as the 
proper Green function with the propagator of external line amputated, i.e.,
\begin{eqnarray}
\langle 0|{\mbox T}\mbox{\boldmath $B$}_A(x)|0\rangle _{prop} & = & \int d^4
y{\it \Delta}_F(x-y)_{B_A}{\mit\Gamma}^{(1)B_A}(y),
\end{eqnarray}
where $\mbox{\boldmath $B$}_A$ stands for the renormalized Heisenberg fields $\mbox{\boldmath
$F$}_\alp,\mbox{\boldmath $F$}^\dag_\alp$ or $\mbox{\boldmath
$D$}_\al$ corresponding to $f_\alp, f^\dag_\alp$ or ${\hat d}_\al$, respectively. 

The momentum space representation ${\hat {\mit\Gamma}}^{(1)B_A}$ of ${\mit\Gamma}^{(1)B_A}(y)$ 
is defined in general by
\begin{equation}
\int d^4 y e^{ipy} {\mit\Gamma}^{(1)B_A}(y) \equiv
(2\pi)^4\delta^4(p){\hat {\mit\Gamma}}^{(1)B_A}(p).\label{eq35}
\end{equation}
Then, as ${\mit\Gamma}^{(1)B_A}(y)$ is written by the propagator ${\it \Delta}_F(y-y)_{ij}=-{\cal M}^{-1}(y-y)_{ij}$
and ${\cal M}$ depends linearly on $f_\alp, f^\dag_\alp$ and ${\hat d}_\al$, we find \cite{Miller}
\begin{equation}
{\hat {\mit\Gamma}}^{(1)B_A} \equiv {\hat {\mit\Gamma}}^{(1)B_A}_{p_{ext}=0} = {\hat
{\mit\Gamma}}^{(1)B_A}(0) = -{\frac12}\frac{\pl}{\pl {\hat
b}_A}\int\frac{d^4k}{(2\pi)^4}\ln\det {\cal M}(k),
\end{equation}
where ${\hat b}_A \equiv \langle B_A\rangle $ and ${\cal M}(k)$ is the momentum representation of ${\cal M}(x)$
(\ref{matrix10}):
\begin{eqnarray}
{\cal M}(k) = \left(\begin{array}{ccc}
		{\cal A}(k) & {\cal B}(k) & 0 \\
		{\cal C}(k) & {\cal M}_{{\hat{\mit\Phi}}{\hat{\mit\Phi}}}(k) & {\cal M}_{{\hat{\mit\Phi}}{\hat A}_5}(k) \\
		0 & {\cal M}_{{\hat A}_5{\hat{\mit\Phi}}}(k) & {\cal M}_{{\hat A}_5{\hat A}_5}(k)
\end{array}\right).
\end{eqnarray}
The estimation proceeds as
\begin{eqnarray}
{\hat {\mit\Gamma}}^{(1)B_A} & = &\!\!\! -{\frac12}\frac{\pl}{\pl {\hat
b}_A}\int\frac{d^4k}{(2\pi)^4}\ln\biggl[\det{\cal M}_{{\hat{\mit\Phi}}{\hat{\mit\Phi}}}(k)\det
\left\{{\cal A}(k) - {\cal B}(k){\cal M}_{{\hat{\mit\Phi}}{\hat{\mit\Phi}}}(k)^{-1}{\cal C}(k)\right\} \nn
& \times & \det{\cal M}_{{\hat A}_5{\hat A}_5}(k)\det\left\{{\cal M}_{{\hat A}_5{\hat A}_5}(k) - {\cal
M}_{{\hat{\mit\Phi}}{\hat A}_5}(k){\cal D}^{-1}(k){\cal M}_{{\hat A}_5{\hat{\mit\Phi}}}(k)\right\}\biggr] \nn
& = &\!\!\! -{\frac12}\frac{\pl}{\pl {\hat b}_A}\int\frac{d^4k}{(2\pi)^4}\ln\det
\left\{{\cal A}(k) - {\cal B}(k){\cal M}_{{\hat{\mit\Phi}}{\hat{\mit\Phi}}}(k)^{-1}{\cal C}(k)\right\}
+ O({\hat g}^5, {\hat g}^4\lambda),
\label{eqa01}
\end{eqnarray}
where
\begin{eqnarray}
{\cal D} = \left(\begin{array}{cc}
		{\cal A}(k) & {\cal B}(k) \\
		{\cal C}(k) & {\cal M}_{{\hat{\mit\Phi}}{\hat{\mit\Phi}}}(k)
\end{array}\right).
\end{eqnarray}

Thus, we obtain
\begin{eqnarray}
&& {\hat {\mit\Gamma}}^{(1)B_A} \nn
& = & -{\frac12}\frac{\pl}{\pl {\hat b}_A}\int\frac{d^4k}{(2\pi)^4}\frac{1}{2l}\ln\det \biggl\{
\left(\begin{array}{cc}
\begin{array}{c} -k^2-\chi^\dag\chi+{\hat g}{\hat d}T \\
-g^2\del(0)(T\eta)(\eta^\dag T) \end{array}              & 
                                                                            -\lambda^\dag f^\dag-g^2\del(0)(T\eta)(T\eta)^t \\
 -\lambda f-g^2\del(0) (\eta^\dag T)^t(\eta^\dag T)    & \begin{array}{c} -k^2-\chi\chi^\dag+{\hat g}{\hat d}T^t
\\
                                                                       -g^2\del(0)(\eta^\dag T)^t(T\eta)^t \end{array}
\end{array}\right) \nn
&& \qquad\qquad + g^2(\del(0)-\half k\coth(lk))\left(\begin{array}{cc}
                           (T\eta)(\eta^\dag T) &  (T\eta)(T\eta)^t  \\
                           (\eta^\dag T)^t(\eta^\dag T)        &  (\eta^\dag T)^t(T\eta)^t
\end{array}\right) + O(g^4)\biggr\} \nn
& \approx & -{\frac12}\frac{\pl}{\pl {\hat b}_A}\int\frac{d^4k}{(2\pi)^4}\frac{1}{2l}
\tr\ln{\it \Delta}({\hat b}), 
\end{eqnarray}
\begin{eqnarray}
{\it \Delta}({\hat b})
& = & [k^2 + \chi^\dag\chi - {\hat g}{\hat d}T+\half g^2k\coth(lk)(T\eta)(\eta^\dag T)]\nn 
&& \qquad\qquad\qquad\q \times [k^2 + \chi\chi^\dag - {\hat g}{\hat d}T^t + \half
g^2k\coth(lk)(\eta^\dag T)^t(T\eta)^t] \nn 
&&  - [\lambda^\dag f^\dag+\half g^2k\coth(lk)(T\eta)(T\eta)^t][\lambda f+\half
g^2k\coth(lk)(\eta^\dag T)^t(\eta^\dag T)],
\label{eqn40}
\end{eqnarray}
where we have performed a Wick rotation and used a formula
\begin{eqnarray}
\sum_{k^5}\frac{(k^5)^2}{k^2 + (k^5)^2} = 2l(\del(0)-\half k\coth(lk)),
\end{eqnarray}
with $k^5$ being summed over the values $\pi n/l(n=\mbox{integer})$, i.e., over whole KK modes. An
interesting observation here is that the $\del(0)$-singularity coming from the KK mode summation has been neatly
cancelled by that from the elimination of the 5D auxiliary field $X^3$. 
\\ 

\noindent{{\bf 5}\q {\bf 1-loop effective potential}}\\

The effective potential $V_{1-loop}$ is nothing but a generator of ${\hat {\mit\Gamma}}^{(1)}(0)$, namely,
\begin{eqnarray}
\frac{\partial V_{1-loop}}{\partial {\hat b}_A} & = & -2l{\hat {\mit\Gamma}}_{P_{ext}=0}^{(1)B_A},\nn
& = & {\frac12}\frac{\pl}{\pl {\hat
b}_A}\int\frac{d^4k}{(2\pi)^4}\ln\det {\cal M}(k), 
\label{eqn:8}
\end{eqnarray}
where ${\hat {\mit\Gamma}}_{P_{ext}=0}^{(1)B_A}={\hat {\mit\Gamma}}^{(1)B_A}(0)$ has
been multiplied by
$2l$ in order for $V_{1-loop}$ to be 4D. The equation (\ref{eqn:8}) is integrated to give
\begin{eqnarray} 
V_{1-loop} = {\frac12}\int\frac{d^4k}{(2\pi)^4}\tr\ln{\it \Delta}(f, f^\dag, {\hat d})+K(\eta,\eta^\dag),
\end{eqnarray}
where $K$ is an integration constant.
 
Finally, we apply the SUSY boundary condition
\begin{equation} 
V_{1-loop}(f = f^\dag = {\hat d}=0)=0, \label{eq58} 
\end{equation}
and obtain 
\begin{eqnarray}
V_{1-loop} & = & {\frac12}\int\frac{d^4k}{(2\pi)^4}\tr[\ln{\it \Delta}(f, f^\dag, {\hat d}) -
\ln{\it \Delta}(0, 0, 0)]\nn
& = & {\frac12}\int\frac{d^4k}{(2\pi)^4}\tr\ln\Bigl[1-\frac{2{\hat g}({\hat d}_\al T^\al)}{k^2 +
\chi^\dag\chi} +\frac{{\cal G}_- - lk\coth(lk){\cal H}}{(k^2 + \chi^\dag\chi)^2} \nn
&& + O(\lambda^4, {\hat g}^4,
{\hat g}^2\lambda^2, {\hat g}^3\lambda)\Bigr],
\label{eqn67}
\end{eqnarray}
where
\begin{eqnarray}
{\cal G}_\pm & \equiv & {\hat g}^2({\hat d}_\al T^\al)^2 \pm (\lambda^{\dag\alp} f^\dag_\alp)(\lambda^\bep
f_\bep),\\  {\cal H} & \equiv & {\hat g}^2\{(\lambda^{\dag\alp}
f^\dag_\alp)(\eta^\dag T^\al)^t(\eta^\dag T^\al)+ (T^\al\eta)(T^\al\eta)^t(\lambda^\alp f_\alp)\}/2  \nn
& + & {\hat g}^3\{({\hat d}_\al T^\al)(\eta^\dag
T^\be)^t(T^\be\eta)^t + (T^\be\eta)(\eta^\dag T^\be)({\hat d}_\al T^\al)\}/2.  
\end{eqnarray}

The resultant effective potential (\ref{eqn67}) is now ready for being integrated w.r.t. the four momentum. Before doing so,
it is useful to comment on the renormalizability. Higher-dimensional field theories are generically non-renormalizable.
The present 5D super-YM model is not exceptional and must be viewed as an effective theory valid up to some high mass
scale associated with an ultraviolet cutoff ${\mit\Lambda}$. However, it should be required that the present model is
renormalizable in the limit of $l \rightarrow 0$.
 
If $T$ has a component such as $\tr T \neq 0$, i.e., the gauge group has a U(1) factor, which we denote as
U(1)$_X$, the term proportional to ${\hat g}T$ in (\ref{eqn67}) provides a dominant contribution, namely
\begin{equation}
V_{1-loop} \approx {\frac12}\int\frac{d^4k}{(2\pi)^4}\tr\ln\left[1-\frac{2{\hat g}{\hat
d}T}{k^2 + \chi^\dag\chi}\right], \label{eq01}
\end{equation}
the integral of which yields a quadratic divergence. As (\ref{eq01}) is independent of $l$, the quadratic divergence
remains in the limit of $l \rightarrow 0$ and will spoil the nonrenormalization theorem. To get around it, we
introduce additional chiral scalar supermultiplets with $\tr Q_X = 0$ in the boundary and require that only one of the
chiral scalar supermultiplets, say
$\phi$, the U(1)$_X$-charge of which is normalized to be 1, has a non-trivial VEV $\eta$. 

The factor $lk\coth(lk)$ in (\ref{eqn67}) alters the high $k$ behaviour of the integrand and yields a term
which is linear in ${\mit\Lambda}$ but suppressed by $l$. In fact, we obtain
\begin{eqnarray}
V_{1-loop} & \approx & -{\frac12}\int\frac{d^4k}{(2\pi)^4}\frac{\tr{\cal G}_++lk\coth(lk)\,\tr{\cal H}}{(k^2 +
\chi^\dag\chi)^2}
\nn & \approx & \frac{1}{16\pi^2}\left[\int_0^{\mit\Lambda}dk\frac{k^3\,\tr{\cal G_+}}{(k^2+\chi^\dag\chi)^2} +
\left\{\int_0^{\tilde k}dk\frac{k^3}{(k^2+\chi^\dag\chi)^2} +
\int_{\tilde k}^{\mit\Lambda}dk\,l\coth(lk)\right\}\tr{\cal H}\right] \nn
& \approx & \frac{1}{32\pi^2}\left[\ln\left(\frac{{\mit\Lambda}^2}{\chi^\dag\chi}\right)\tr{\cal
G}_+ - l{\mit\Lambda}\,\tr{\cal H}\right] + \mbox{finite part}, \label{eq92}
\end{eqnarray}
where terms that vanish as ${\mit\Lambda}$ goes to infinity have been neglected and we have split the integral
of the term prpportional to $\tr{\cal H}$ into two regions, $0 \leq k \leq {\tilde k}$ and ${\tilde k} \leq
k \leq {\mit\Lambda}$, assuming $\vert \chi \vert \ll {\tilde k} \ll l^{-1}$. 

The leading term in r.h.s. of (\ref{eq92}) is apparently linealy divergent. However, the cutoff ${\mit\Lambda}$ is
multiplied by the length $l$ of the extra dimension which may suppress the growth of ${\mit\Lambda}$ so as to give a
finite and significant contribution to the effective potential. This is nothing but a bulk effect and plays an important role
in the minimization of the effective potential as will be described in the next section.\\

\noindent{{\bf 6}\q {\bf Minimization of the effective potential}}\\

The utility of effective potential is to search a true vacuum by its minimization. Our effective potential is,
however, SUSY so that it vanishes trivially at the minimum point $f = {\hat d} = 0$. In order to examine its
physical property, therefore, it is appropriate to introduce a term such as FI
$D$-term\footnote{Such a $D$ term has been introduced into the
hidden sector in \cite{Mirab}.} into the boundary Lagrangian ${\cal L}_{bnd}$ and observe how the spontaneous
breaking of SUSY as well as that of gauge symmetry is realized. For this purpose, we choose the gauge group to be
U(1)$_X$ discussed in the previous section.

The effective potential to be minimized is then as follows:
\begin{eqnarray}
V^{eff} = && \!\!\!\!\!\!\!\!\!\! V_{tree} + V_{1-loop} + V_{FI},\\
V_{tree} & = & - f^\dag f + \eta^t mf + \eta^\dag mf^* + \half\eta^t\lambda f\eta +
\half\eta^\dag\lambda^\dag f^\dag\eta^* - \half{\hat d}^2 - {\hat g}\eta^\dag {\hat d}\eta, \\
V_{1-loop} & = & - \al \tr\{(\lambda^\dag f^\dag\eta^{\dag 2} + \eta^2\lambda f)/2
 + {\hat g}\eta^\dag{\hat d}\eta)\},\\
V_{FI} & = & -\xi {\hat d},
\end{eqnarray}
where $V_{tree}$ is a tree level potential which is directly read from (\ref{mp2}) and (\ref{mp7}),
$V_{1-loop}$ is the dominant part of (\ref{eq92}) with
\begin{eqnarray}
\al \equiv \frac{l{\mit\Lambda}{\hat g}^2}{32\pi^2},
\end{eqnarray}
and $V_{FI}$ comes from the FI $D$-term ${\cal L}_D = \xi D^{(5)}$.

In order to trace essential features of our analysis, we assume that the components of each
classical scalar field vanish except for a certain real component which we denote by the same symbol. 
Then, we have
\begin{eqnarray}
V^{eff}(f,{\hat d},\eta) = -f^2 + 2mf\eta + \lambda f (1-\al)\eta^2 - \half{\hat d}^2 - (1+\al){\hat
g}{\hat d}\eta^2 - \xi {\hat d}.
\end{eqnarray}
The auxiliary fields $f, {\hat d}$ are written as functions of $\eta$ through the conditions $\pl V^{eff}/\pl f =
\pl V^{eff}/\pl {\hat d} = 0$ as follows;
\begin{eqnarray}
f & = & m\eta + \frac{\lambda}{2}(1-\al)\eta^2 \equiv {\tilde f},\\
{\hat d} & = & - \xi - (1+\al){\hat g}\eta^2 \equiv {\tilde d},\\
\end{eqnarray}
by which we eliminate $f, {\hat d}$ from $V^{eff}$:
\begin{eqnarray}
V^{eff}({\tilde f},{\tilde d},\eta) & = & V^{eff}(\eta) = \left\{\frac{\lambda^2(1-\al)^2}{4}
+ \frac{(1+\al)^2{\hat g}^2}{2}\right\}\eta^4 +\lambda m(1-\al)\eta^3 \nn
&& + \left\{m^2+(1+\al){\hat g}\xi\right\}\eta^2 +\frac{\xi^2}{2}. \label{mp13}
\end{eqnarray}
From now on, we assume that $\phi$ is massless ($m = 0$ ) for simplicity. Then, if ${\hat g}\xi > 
0, V^{eff}(\eta)$ has a minimum at $\eta = 0$ with a minimum value $\xi^2/\{2(1+2\al)\}$, which measures
the SUSY breaking scale $M_{SUSY}$. The gauge symmetry is not broken. If ${\hat g}\xi < 0$, on the other
hand, $V^{eff}(\eta)$ is minimized at
\begin{eqnarray}
\eta^2 & = & \frac{-2(1+\al){\hat g}\xi}{\lambda^2(1-\al)^2 + 2(1+\al)^2{\hat g}^2} \equiv {\tilde\eta}^2,
\end{eqnarray}
with the minimum value
\begin{eqnarray}
V^{eff}(\eta={\tilde \eta}) = {\tilde V}^{eff} =
\frac{(1-\al)^2\lambda^2}{(1-\al)^2\lambda^2 +2(1+\al)^2{\hat g}^2}\frac{\xi^2}{2},
\label{mp10}
\end{eqnarray}
which is a functiom of $\al$ for given $\lambda, {\hat g}$ and $\xi$ and has an absolute minimum at 
$\al = 1$ where ${\tilde V}^{eff} = 0$ as shown in Fig.1. 
\begin{figure}
  \begin{center}
    \includegraphics[width=130.5mm,height=88.55mm]{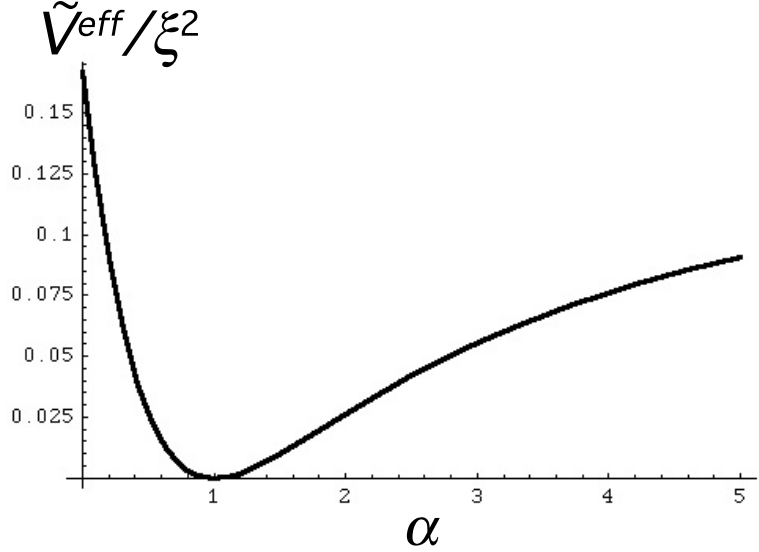}
  \end{center}
  \caption{$\al$-dependence of ${\tilde V}^{eff}$ for $\lambda = {\hat g}$. }
  \label{Fig1}
\end{figure}
Therefore, the size of the extra dimension is settled
at $l^{-1}={\mit\Lambda}{\hat g}^2/32\pi^2$ making SUSY restored in the true
vacuum. For example, $l^{-1} \approx 3.0 \times10^{11}$ GeV ($l^{-1} = 7.2 \times10^{13}$ GeV) for ${\mit\Lambda}
= M_{GUT} \approx 10^{16}$ GeV (${\mit\Lambda} = M_{Pl} = 2.4 \times10^{18}$ GeV) and ${\hat g} \approx 0.1$. 

At $\al = 1$, (\ref{mp13}) becomes
\begin{eqnarray}
V^{eff}(\eta) = 2\left({\hat g}\eta^2 + \frac{\xi}{2}\right)^2, \label{mp14}
\end{eqnarray} 
for any $\lambda$, which has minima at $\eta = \pm\sqrt{\displaystyle{\bigl |\xi/2{\hat g}}\bigr |}
\equiv {\tilde \eta}$ (hence ${\tilde f} = {\tilde d} = 0$) as shown in Fig.2, provided ${\hat g}\xi < 0$. Namely
$\phi$ plays a role of Higgs field which breaks the gauge 
\begin{figure}
  \begin{center}
    \includegraphics[width=312pt,height=212.5pt]{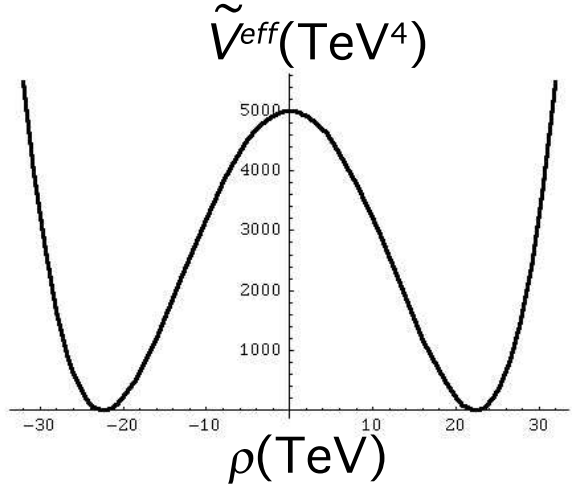}
  \end{center}
  \caption{$\eta$-dependence of $V^{eff}$ for ${\hat g}=0.1$ and $\xi = 100 \mbox{TeV}^2$.}
  \label{Fig2}
\end{figure}
symmetry with the breaking scale  $\langle\phi\rangle  = {\tilde \eta}$, while SUSY is restored in spite of the presence
of FI $D$-term.\footnote{Such a phenomenon is known to occur for $\al = 0$, i.e., at the tree level, too, only if
$\lambda = 0$ as far as ${\hat g}\xi < 0$. In our case ($\al = 1$), (\ref{mp13}) is valid irrespectively of
$\lambda$.} It is not $M_{SUSY}$ but the gauge symmetry breaking scale ${\tilde \eta}$ that $\xi$ affects. 
\\

\noindent{{\bf 7}\q {\bf Concluding remarks}} \\

We have estimated a SUSY effective potential of the 5D super-YM model with the extra
dimension compactified on $S^1/Z_2$ at the 1-loop level. Under such assumptions that the quadratic
divergence does not arise, its dominant part is apparently linealy divergent and proportional to $l{\mit \Lambda}$, i.e., a
product of the size of extra dimension and the cutoff scale. If $l$ is small but much bigger than the "cutoff
size" ${\mit \Lambda}^{-1}$ corresponding to the Planck, string or GUT scale, the divergence is suppressed
and the term proportional to $l{\mit\Lambda}$ of 1-loop effective potential proves to be finite and not negligible.
This is just the bulk effect which originates from taking in all the KK excitation modes of the bulk
field $\mit\Phi$ and reveals an interesting situation. In fact, taking the tree level contributions
and the FI $D$-term into account, we find that the effective potential is minimized at a specific value of $l$,
where SUSY is restored but the gauge symmetry is broken. It is remarkable that the value of $l$ corresponds to an
intermediate energy scale where new ingredients of gauge theory are expected to be disclosed. 

As an approach to regard the extra-space radius as a dynamical variable, the radion model is, at present,
most promising. There, the radius parameter $l$ is regarded as a vacuum expectation value of the field "radion" \cite{{RS}, {GW}}.
It would be more complete to treat the result of Sec.6 in the framework of radion model. In order to incorporate the radion 
and the dilaton in a multiplet, we are naturally led to consider 5D supergravity (SUGRA). Consistency of FI-terms in the 5D 
SUGRA has been investigated in \cite{BCCRS}. It remains as a future work to examine the conclusion of our model in 
connection with the radius stabilization in this context.

Such a phenomenon that the radiative correction appears to be proportional to $l{\mit\Lambda}$ seems inherent in gauge 
theories with extra dimensions compactified on flat space. Indeed, it has
been observed that the renormalization group running of the gauge coupling constants changes from "logarithmic" to
"linear" in a 5D version of minimal SUSY standard model with the flat extra dimension compactified on
$S^1/Z_2$  \cite{Dienes}. This fact is due to the presence of infinite towers of KK states and causes an accelerated
unification of strong, electromagnetic and weak couplings only a little above $\mu_0 \equiv l^{-1}$. However, if the
extra 5th dimension is warped as in the case of Randall-Sundrum (RS) \cite{RS}, the gauge coupling running can be
logarithmic \cite{{Pom},{RSchw}}. It is, therefore, worth to try to compute the SUSY effective potential in
the 5D super-YM model with the RS background and examine whether it is minimized at a non-trivial value of
the radius of extra dimension or not. The bulk effect to our SUSY effective potential is principally due to the contribution
from the bulk propagator of $\mit\Phi$. Since the interaction of $\mit\Phi$ with $\phi$'s takes place only in the 4D
boundary, the relevant loop amplitude including the the $\mit\Phi$-propagator might not be affected by $x^5$-dependent
cutoff \cite{RSchw} in the RS background, so that the "linear" growth of effective potential along with the cutoff would
have a popssibility to be retained even if the extra dimension is warped. The details will be discussed in the forthcoming
paper \cite{Ichinose2}.\\

\noindent{{\bf Acknowledgment} \\

This work originates in discussions at Chubu summer school 2003. The authors would like to thank the participants and
the Yukawa Institute for Theoretical Physics which supported the school. 

\end{document}